\documentstyle[12pt,aasms4]{article}
\received{}
\revised{}
\accepted{}

\cpright{}{}
\journalid{}{}
\articleid{}{}
\paperid{}

\cpright{}{}
\ccc{}

\lefthead{Grandpierre}
\righthead{ }

\begin{document}

\title{ A Dynamic Solar Core Model: the Deviant Temperatures Approach }

\author{Attila Grandpierre}

\affil{Konkoly Observatory, P.O. ~Box ~67, H--1525,
Budapest, Hungary \\
electronic mail: grandp@ogyalla.konkoly.hu}

\authoraddr{Konkoly Observatory, P.O. ~Box ~67, H--1525,
Budapest, Hungary}

\vskip 10cm

\begin{abstract}
I derive here a model independent inequality which shows that the SuperKamiokande contains a term arising from a non-pp,CNO source. 
First principle physics shows that the non-pp,CNO source is of thermonuclear runaway origin. Several indications suggest strongly that the non-pp,CNO term plays a more significant role in the solar neutrino problems than neutrino oscillations. When removing the over-restricted standard solar model luminosity constraint, the temperature dependence of the neutrino fluxes is related to pure nuclear physics and follows $\Phi_{pp} \propto T^4$ instead of $\Phi_{pp} \propto T^{-1/2}$. The results of the calculations offer a new way to solve the solar neutrino problems and problems of neutrino oscillations. The dynamic solar core model presents predictions to Borexino and SNO measurements. These predictions can serve to distinguish between the MSW and the non-pp,CNO effect.  

{\it PACS numbers}: 26.65+t, 26.30.+k, 96.60Jw, 95.30.Cq

\end{abstract}

\section{Introduction}

	The first experimental checks of the hypothesis that the Sun 
produces its energy mainly through the proton-proton cycle indicated 
that it may not be correct since the (three) neutrino problems [1] arose. I point out that the solar neutrino problems - together with the atmospheric neutrino anomaly [2] and the anomalous results of the Liquid Scintillator Neutrino Detector (LSND), when allowing three neutrino flavours, are not consistent [3]. I argue that a fundamental thermonuclear instability, revealed by first principles of physics by Grandpierre [4]-[6] and Zel'dovich, Blinnikov and Sakura [7], offers a way to make the neutrino problems consistent. A detailed introduction of the problem will follow in [8].

%
%
\section{ Basic equations: the deviant temperatures approach }
\begin{eqnarray}
S_K = a_{K8} \Phi_8
\end{eqnarray}
\begin{eqnarray}
S_C = a_{C1} \Phi_1 + a_{C7} \Phi_7 + a_{C8} \Phi_8
\end{eqnarray}
\begin{eqnarray}
S_G = a_{G1} \Phi_1 + a_{G7} \Phi_7 + a_{G9} \Phi_8 ,
\end{eqnarray}
with a notation similar to that of Heeger and Robertson [9]: the subscripts i = 1, 7 and 8 refer to $pp + pep$, $Be + CNO$ and $B$ reactions. The $S_j$-s are the observed neutrino fluxes at the different neutrino detectors, in dimensionless units, j = K, C, G to the SuperKamiokande, chlorine, and gallium detectors. $\Phi_i$ are measured in $10^{10} \nu cm^{-2}s^{-1}$. Similar equations are presented by Castellani et al. [10], Calabresu et al. [11], and Dar and Shaviv [12] with slightly different parameter values.
Using these three detector-equations to determine the individual neutrino fluxes  $\Phi_i$, I derived that
\begin{eqnarray}
\Phi_8 = S_K/a_{K8}
\end{eqnarray}
\begin{eqnarray}
\Phi_1 = (a_{G7}S_C -a_{C7}S_G + S_K/a_{K8}(a_{C7}a_{G8} - a_{G7}a_{C8}))/D
\end{eqnarray}
\begin{eqnarray}
D = a_{G7}a_{C1} - a_{C7}a_{G1}
\end{eqnarray}
and
\begin{eqnarray}
\Phi_7 = (a_{G1}S_C - a_{C1}S_G + S_K/a_{K8}(a_{C1}a_{G8}-a_{G1}a_{C8}))/D'
\end{eqnarray}
\begin{eqnarray}
D'= a_{G1}a_{C7} - a_{C1}a_{G7}.
\end{eqnarray}
\begin{eqnarray}
\Phi_7 \times D' = a_{G1}S_C - a_{C1}S_G + (a_{C1} a_{G8}/a_{K8} - a_{G1} a_{C8}/a_{K8})S_K.
\end{eqnarray}
Obtaining these solutions, the root of the beryllium-problem goes to the circumstance that $D'> 0$, therefore the numerator has to be also positive. We know that $\Phi_7$ can have only a physical, positive value. This fact requires that the following formulae has to be valid: 
\begin{eqnarray}
S_K < (a_{G1}S_C-a_{C1}S_G)/(a_{C1}a_{G8}/a_{K8}-a_{G1}a_{C8}/a_{K8})
\end{eqnarray}
Numerically,
\begin{eqnarray}
\Phi_7 = 0.4647 S_C -0.0014S_G - 0.5125S_K > 0.
\end{eqnarray}
Now we see that the problem of the suppression of solar beryllium neutrinos is related to the circumstance  
that the coefficient of $S_C$ is smaller than  $S_K$, therefore, since all the $S_i$ values are positive, $\Phi_7$ cannot be physical. If we require a physical $\Phi_7$, with the numerical values of the detector sensitivity coefficients, 
this constraint will take the following form:
\begin{eqnarray}
S_K < 0.9024S_C - 0.0027S_G \simeq 2.115
\end{eqnarray}
with the observed values $S_K = 2.44$, $S_C = 2.56$, and $S_G = 72.2$ [11]. 
This inequality of the neutrino detector rates (10) was not derived previously in the published literature by my knowledge. This is a completely model independent inequality which shows that the SuperKamiokande (and possibly the other detectors as well) contains a term arising from neutrinos of a non-pp,CNO source. That is, with standard neutrinos, it is not physical to use the SuperKamiokande result $S_K$ at its face value $2.44$ in the standard neutrino equations. I have shown here that this fact is the basic root of the problem of the missing beryllium neutrinos. The detector rate inequality can be fulfilled only if we introduce an additional term $S_K(x)$ to represent the contribution of non-pp,CNO neutrinos to the (Super)Kamiokande measurements (and, in principle, if we modify the $S_C$, $S_K$ values properly). The presence of a non-electron (non-SSM) neutrino term in the SuperKamiokande is interpreted until know as indication to neutrino oscillations. Nevertheless, thermal runaways are indicated to be present in the solar core producing high-energy electron neutrinos, as well as, possibly, muon and tau neutrinos. Moreover, the explosive reactions have to produce high-energy axions to which also only the SuperKamiokande is sensitive. These indications suggest a possibility to interpret the neutrino data with standard neutrinos as well. 

What information can be subtracted from the neutrino flux equations about this extra term $S_K(x)$? To see this, I introduced our "a priori" nuclear physics knowledge on the pp,CNO chains, i.e. their temperature dependence. In this way one can derive the different temperatures in the solar core at different characteristic depths belonging to the $pp$, $Be$ and $B$ neutrino productions. I note that the introduction of temperature dependence does not lead to solar model dependency. Instead, it points out the still remaining solar model dependencies of the previous SSM calculations and allowing other types of chains, it removes a hypothetical limitation. Accepting the presence of explosive chains as well, it probably presents a better approach to the actual Sun.

I allow different effective temperatures at the different depths characteristic for the beryllium and CNO neutrinos, $T_{int}=T_{Be+CNO}$, boron neutrinos ($T_B$) and proton-proton neutrinos ($T_p=T_{pp+pep}$), as it is suggested for the astrophysical solution of the solar neutrino problem(s) [12]-[15]. Assuming that $T=T_p \simeq T_{int}$, I will have three equations for the three unknown variables $T$, $T_B$ and the non-pp,CNO part of the Super-Kamiokande data $S_K(x)$. In this way, I can deduce the value $S_K(x)$ and also $T_B$ from the temperatures present in the helioseimically better known, outer parts of the Sun. 

An essential point in my calculations is that I have to use the temperature dependence proper in the case when the luminosity is not constrained by the SSM luminosity constraint, because another type of energy source is also present. The SSM luminosity constraint and the resulting composition and density readjustments, together with the radial extension of the different sources of neutrinos modify this temperature dependence. The largest effect arises in the temperature dependence of the $pp$ flux: $\Phi_1 \propto T^{-1/2} $ for the SSM luminosity constraint (see the results of the Monte-Carlo simulations of Bahcall and Ulrich [16]), but $\Phi_1 \propto T^4$ without the SSM luminosity constraint. Inserting the temperature-dependence of the individual neutrino fluxes for the case when the solar luminosity is not constrained by the usual assumption behind the SSM [12] into the chlorine-equation, we got the temperature dependent chlorine equation
\begin{eqnarray}
S_C(T) = a_{C1}T^4 \Phi_1(SSM) + a_{C7}T^{11.5} \Phi_7(SSM) + a_{C8}T_B^{24.5} \Phi_8(SSM)	
\end{eqnarray}
Similarly, the temperature-dependent gallium-equation will take the form:
\begin{eqnarray}
S_G(T) = a_{G1}T^4 \Phi_1(SSM)+ a_{G7}T^{11.5} \Phi_7(SSM) + a_{G8}T_B^{24.5} \Phi_8(SSM), 
\end{eqnarray}
and the SuperKamiokande equation shows that
\begin{eqnarray}
S_K(T)=a_{K8}T_B^{24.5} \Phi_B(SSM) + S_K(x).
\end{eqnarray}
Eliminating $T_B$ from (13) and (14), an equation is obtained for $T$,
\begin{eqnarray}
T^{11.5} \Phi_{int}(SSM)(a_{G8}a_{C7} - a_{C8}a_{G7}) + T^4\Phi_p(SSM)(a_{C1}a_{G8} - a_{C8}a_{G1}) = S_Ca_{G8} - S_Ga_{C8}
\end{eqnarray}

With $\Phi_{pp}(SSM) = 5.94 \times 10^{10} cm^{-2}s^{-1}$, $\Phi_{Be}(SSM)=4.80 \times 10^9 cm^{-2}s^{-1}$, $\Phi_B(SSM)=5.15 \times 10^6 cm^{-2}s^{-1}$, $\Phi_B(SK,obs)=2.44 \times 10^6 cm^{-2}s^{-1}$, $S_C=2.56$, $S_G=72.2$ from [11], $\Phi_p(SSM) = 5.95 \times 10^{10} cm^{-2}s^{-1}$, $\Phi_{int}(SSM) = 0.594 \times 10^{10} cm^{-2}s^{-1}$ and $\Phi_B(SSM) = 0.000515 \times 10^{10} cm^{-2}s^{-1}$. 

With these values, equation (16) will take the form
\begin{eqnarray}
T^{11.5} + 1.713 T^4 = 1.572,
\end{eqnarray}
the solution of which will give a value of $T \simeq 0.916$. With this $T$, $T_B$ will be $T_B \simeq 0.956$, $S_K(x) = 0.81 \times 10^6 cm^{-2}s^{-1}$, 
$\Delta R_K(x) = R_K(obs)-R_K(T=0.956) = \Phi_K(obs)/\Phi_K(SSM) - \Phi_B(T=0.956)/\Phi_B(SSM) \simeq 0.142$, which is around $30 \%$. 

%
%
\section{Discussion}
The results obtained above suggest that a yet unrecognised class of astrophysical solutions to the solar neutrino problem(s) may be at work in the Sun. 
The key element of this solution is to allow the presence of a non-pp,CNO energy source to be active in the Sun. The presence of a non-pp,CNO energy source modifies the SSM luminosity constraint, and through this circumstance also the temperature dependence of the individual neutrino fluxes. 
This non-pp,CNO energy source produces additional neutrino flux at the SuperKamiokande. We know that the SuperKamiokande is the only detector, which is sensitive to neutral currents, fluxes of muon and tau neutrinos. If we ignore at present the yet hypothetical MSW effect to be at work here, we need a source that is able to generate muon (tau) neutrinos in the Sun. To generate muon neutrinos, it is necessary a high temperature (the estimated temperature is  $ \leq 10^{11} K$), and high density for a significant amount of muon-neutrino flux. This temperature is just the one that is calculated for the hot bubbles [6]. Moreover, the hot bubbles are able to generate high-energy axions, and only the SuperKamiokande detector is sensitive to detect axions [19]. The contribution of the anti-neutrinos may be present also only in the SuperKamiokande data, but it seems not to be significant, as being less than $5.2 \%$ [20].

In the approach of "deviant temperatures" I derived new type of equations and from it I determined the proton-beryllium temperature $T=0.916$ and the boron temperature $T_B=0.956$. This result suggest that the Sun is relatively hotter in the innermost $5 \%$ solar radius. On the other hand, a solution to the solar neutrino problem is obtained. The beryllium neutrino flux in the dynamic solar model (DSM) is $\Phi_{Be}(DSM) = T^{11.5} \Phi_{Be}(SSM) = 1.75 \times 10^9 cm^{-2}s^{-1}$. 

The result obtained by the above calculations gives expectation values for the future neutrino measurements different from the MSW calculations. For example, the dynamic solar model (DSM) suggest a $R_{Be} = \Phi_{Be}(observed)/\Phi_{Be}(SSM) \leq 36 \%$ relative depletion of the beryllium-neutrino flux when compared to the expected SSM value, and $R_B = \Phi_B(T=0.956)/\Phi_B(SSM) \leq 33 \%$ depletion for the boron-neutrino flux. These relative depletions refers to the "quiet solar core", to the part of the core without the runaway "hot bubbles" regions. Of course, the non-pp,CNO runaway term produce an additional increase in the high energy neutrino spectrum. 

The differences between these depletions as obtained here differ strongly from the ones of the small-angle MSW solution, where $R_{Be} \simeq 0$ and $R_B \simeq 40 \%$ ([21]-[24]). On the other hand, the large-angle MSW solution shows a nearly constant depletion above $1MeV$, $R(>1MeV) \simeq 0.2$ (see Fig. 16, in [25]). Therefore, the large angle solution of the MSW effect is not consistent with the preferential high-energy enhancement of the neutrino spectra ([9], [26], [27]). If the mechanism suggested here - the functioning of a new nuclear reaction channel - works in the real Sun, then the Borexino has to measure a $ \Phi_b< R_{Be}(DSM) < 0.36 + \Phi_b$, for $0 < T < 0.916$. The derived results suggest values close to the upper limit. With T=0.916, the DSM prediction to Borexino is $\Phi_{Be}(DSM) = \Phi_{Be}(SSM) \times 0.36 + \Phi_b$ (here $\Phi_b$ is the bubble- and/or the non-pp,CNO neutrino flux), while the small-mixing-angle MSW solution would suggest a value close to zero. In this way, the derived predictions offer a possibility to distinguish between the case in which the MSW effect dominates $\Phi_{Be}(obs.) \simeq 0$, the other case in which a hybrid MSW+DSM mechanism works $0 < \Phi_{Be}(obs.) < 0.36$ and the third case in which the DSM mechanism works alone $\Phi_{Be}(obs.) \simeq 0.36$, or larger. Moreover, the spectrum above $5MeV$ should show a significant enhancement towards the larger energies produced by the bubble-term. The dynamic solar model allows a larger place for $\Phi_{Be}$ because it works with a lower value for the standard neutrino fluxes at Kamiokande. Lowering the $pp$ neutrino flux, more place remains to the $Be$ neutrinos as well. Regarding the future measurements of SNO, the prediction of DSM is $\Phi_{\nu_e}(SNO, DSM) \simeq 0.33 \times \Phi_{\nu_e}(SNO,SSM) + \Phi_{\nu_e}(bubbles)$. Therefore, the charged current $CC(DSM) \simeq CC(SSM) \times 0.33 + CC(bubbles)$, and the neutral currents $NC(DSM) \simeq 1/3 \ CC(SSM) \times 0.33 + NC(bubbles)$.
 
If the pp-temperature is $T \simeq 0.916 T(SSM)$, this means that the pp luminosity of the Sun is only $L_{pp} \simeq 70 \% L(SSM)$. The remaining part of the solar luminosity should be produced by the hot bubbles, 
$L_b \simeq 30 \% L(SSM)$. The new type nuclear reactions proceeding in the bubbles (and possibly in the microinstabilities) should also produce neutrinos, and this additional neutrino-production, $\Phi_b$ should generate the surplus terms in the chlorine and water Cherenkov detectors as well. At present, I was not able to determine which reactions would proceed in the bubbles, and so it is not possible to determine the accompanying neutrino production as well. Nevertheless, it is plausible that at that high temperature such nuclear reactions occur as at nova-explosions or other types of stellar explosions. Admittedly, these could be rapid hydrogen-burning reactions, explosive CNO cycle, and also nuclear reactions producing heat but not neutrinos, like e.g. the explosive triple-alpha reaction ([28]-[30]). At present, I remark that the 
calculated bubble luminosity ($ \simeq 30 \%$) may be easily consistent with the calculated non-pp,CNO neutrino flux $R_K(x) \simeq 30 \%$.

The relation between the non-pp,CNO neutrino fluxes and the non-pp,CNO luminosity, together with the relation between the pp,CNO neutrino fluxes and the relevant pp,CNO luminosity, would stand to the place of the over-restricted solar luminosity constraint (this would be the generalised luminosity constraint). The above results are in complete agreement with the conclusion of Hata, Bludman and Langacker [22], namely: "We conclude that at least one of our original assumptions are wrong, either (1) Some mechanism other than the $pp$ and the $CNO$ chains generates the solar luminosity, or the Sun is not in quasi-static equilibrium, (2) The neutrino energy spectrum is distorted by some mechanism such as the MSW effect; (3) Either the Kamiokande or Homestake result is grossly wrong." These conclusions are concretised here to the following statements: (1) a non-pp,CNO energy source is present in the solar core, and the Sun is not in a thermodynamic equilibrium, (2) this non-pp,CNO source distorts the standard neutrino energy spectrum, and perhaps the MSW effect also contributes to the spectrum distortion (3) The SuperKamiokande  results (and, in principle, the Homestake, Gallex also) contains a term arising from the non-pp,CNO source. The Homestake and Gallex may observe the high-energy electron neutrinos produced in the hot bubbles. 

The results presented here suggest that the beryllium neutrino flux is lower than expected by the SSM because the neutrino temperatures (as measured by the different neutrino detectors) are lower than the expected SSM-value. The main reason is that because a non-pp,CNO energy source is also present in the solar core, the quiet SSM-like solar core may have a lower neutrino temperature. Nevertheless, if the SSM electron neutrinos do take part in neutrino oscillation, the oscillation would lead to another factor which would depress the intermediate energy neutrinos, besides the apparent "cooling". 

One may think that the suggested mechanism could solve the solar neutrino problems, but new problems arose: the problem of the apparent "cooling" of the solar core, and the problem how the dynamic Sun is consistent with the helioseismic measurements. Although these questions would lead to another field, which does not necessarily belong to the present topic, let me outline some preliminary considerations. One thing is that the seismic temperatures and the neutrino temperatures do not necessarily wear the same values. The presence of an explosive energy source decouples the neutrino fluxes and temperatures from the
seismic temperatures. Gavryusev [31] pointed out, that it is not possible to deduce directly the central temperature from the solar seismological data. Solar model calculations did show that the sound speed is an average property of the whole star and cannot be connected in any way to an "average temperature". The sound speed, as deduced from solar oscillations, is an "averaged sound speed" and it is a very stable value defined by global solar parameters (mass, radius, luminosity). Even significant changes in the inner solar model structure do not change it much. 

On the other hand, we can pay attention to the fact that the energy produced in the solar core do not necessarily pours into thermal energy, as other, non-thermal forms of energy may also be produced, like e.g. energy of magnetic fields. The production of magnetic fields can significantly compensate the change in the sound speed related to the lower temperature, as the presence of magnetic fields may accelerate the propagation of sound waves with the inclusion of magnetosonic and Alfven magnetohydrodynamical waves.  Moreover, it seems that one need more careful analysis and more physical inputs to interpret properly the helioseismic data in the study of the innermost structure of the solar core in detail, since it is sensitive to the complex conditions present in the solar core. These include not only to the factors included in the standard solar model, but also to magnetic fields, nonequilibrium thermodynamics, and all the different manifestations of the energy production occurring in the solar core. And since together these sum up to the solar luminosity, the helioseismic data may allow cooler than standard Sun as well, because the thermal energy contains only a part of the total produced energy.  

The continuously present microinstabilities should produce a temperature distribution with a double character, as part of ions may posses higher energies. Their densities may be much lower than the respective ions closer to the standard thermodynamic equilibrium, and so they may affect and compensate the sound speed in a subtle way. Recent calculations of the non-maxwellian character of the energy distribution of particles in the solar core ([32], and more references therein) indicate that the non-maxwellian character leads to lowering the SSM neutrino fluxes and, at the same time, produces higher central temperatures. This effect may also compensate for the lowering of the sound speed by the lowering of central temperature. 

At the same time, an approach specially developed using helioseismic data input instead of the luminosity constraint, the seismic solar model indicates a most likely solar luminosity around $0.8 L_{Sun}$([33], Figs. 7-10), which leads to a seismological temperature lower than its SSM counterpart,  $\Delta T \simeq 6 \%$. 
On the other hand, as Bludman et al. [34] pointed out, the production of high energy $^8B$ neutrinos and intermediate energy $^7Be$ neutrinos depends very sensitively on the solar temperature in the innermost $5 \%$ of the Sun's radius. In the region below $0.2$ solar radius the actual helioseismic datasets do not seem to offer reliable results ([35], see also [36], [37]).

%
%
\section{Conclusions}
I derived a new, really model independent inequality which shows that a neutrino flux is present besides the standard pp,CNO sources (at least) in the SuperKamiokande data. It is shown that first principle physics suggests that thermonuclear runaways produce non-pp,CNO nuclear reactions [8]. The runaway source produces high-energy axions, and may easily produce high-energy electron, muon and tau neutrinos as well. The contribution of this non-SSM source to the SuperKamiokande data is estimated. The non-pp,CNO photon and neutrino luminosity seem to be consistent. Moreover, the relative temperature of the innermost solar core is found to be higher than farther from the centre. Predictions of the dynamic solar core model are presented for the Borexino and SNO measurements that can distinguish between the cases when the MSW effect is dominant, the hybrid MSW+DSM mechanisms works, or the DSM mechanism dominates. 

The solution of the model independent neutrino flux equations strongly suggests that a new type of energy production mechanism is present in the solar core. 
The non-pp,CNO reactions are suggested to contribute to the production of intermediate, and, preferentially, of high energy neutrinos. Therefore, they are able to distort the solar neutrino spectrum in the way as it is indicated in [9], [27], without invoking new neutrino physics. The higher depletion of the intermediate energy neutrinos arises as a consequence of the lower than standard neutrino temperatures. The preferential enhancement in the high-energy region of the neutrino spectra is interpreted as enhanced by the contribution of thermonuclear runaways produced in micro- and macro-instabilities. 

The indicated presence of a non-pp,CNO energy source in the solar core - if it will be confirmed - will have a huge significance in our understanding of the Sun, the stars, and the neutrinos. This subtle and compact phenomena turns the Sun from a simple gaseous mass being in hydrostatic balance to a complex and dynamic system being far from the thermodynamic equilibrium. This complex, dynamic Sun ceases to be a closed system, because its energy production is partly regulated by tiny outer influences like planetary tides. This subtle dynamics is possibly related to stellar activity and variability. Modifying the participation of the MSW effect in the solar neutrino problem, the dynamic energy source has a role in the physics of neutrino mass and oscillation.
An achievement of the suggested dynamic solar model is that it may help to solve the physical and astrophysical neutrino problems without the introductio n of sterile neutrinos, and, possibly, it may improve the bad fit of the MSW effect[38].

%
%
\section{Acknowledgements}
The work is supported by the Hungarian Scientific Research Foundation
OTKA under No. T 014224.

\eject

\end{document}